\documentclass[11pt,twoside]{article}
\usepackage{asp2010}
\usepackage{graphics}
\newcommand{\gta}{\lower 0.5ex\hbox{$ \buildrel>\over\sim\ $}}
\newcommand{\lta}{\lower 0.5ex\hbox{$ \buildrel<\over\sim\ $}}

\begin{document}
\title{Observational Asteroseismology of Hot Subdwarf Stars with the
  Mont4K/Kuiper Combination at the Steward Observatory Mount Bigelow Station}
\author{G. Fontaine$^{1}$, E.M. Green$^{2}$, S. Charpinet$^{3}$,
  M. Latour$^{1}$, S.K. Randall$^{4}$, V. Van Grootel$^{5}$,
  P. Brassard$^{1}$, and several undergraduate students at University of
  Arizona} 
\affil{$^{1}$D\'epartement de Physique, Universit\'e de Montr\'eal,
  Montr\'eal, Qu\'ebec, Canada H3C 3J7\\
       $^{2}$Steward Observatory, University of Arizona, 933 North
  Cherry Avenue, Tucson, AZ 85721, USA\\ 
       $^{3}$CNRS, Universit\'e de Toulouse, UPS-OMP, IRAP, 14
  av. E. Belin, 31400 Toulouse, France\\
       $^{4}$European Southern Observatory, Karl-Schwarzschild-Str. 2,
  85748 Garching bei M\"unchen, Germany\\
       $^{5}$Institut d'Astrophysique et de G\'eophysique de
  l'Universit\'e de Li\`ege, All\'ee du 6 Ao\^ut 17, B-4000 Li\`ege,
  Belgique}

\begin{abstract} 
In the last few years, we have carried out several extensive
observational campaigns on pulsating hot subdwarf stars using the Mont4K
CCD camera attached to the 1.55 m Kuiper Telescope on Mount Bigelow. The
Mont4K is a joint partnership between the University of Arizona and
Universit\'e de Montr\'eal. It was designed and built at Steward
Observatory. Using the Mont4K/Kuiper combination, we have so far, and
among others, gathered high-sensitivity broadband light curves for PG
1219+534, PB8783, HS 0702+6043, and Feige 48. We report very briefly on
some of the most interesting observational results that came out of
these campaigns.
\end{abstract}

\section{The Mont4K Instrument}

The Mont4K (as in Montr\'eal 4K $\times$ 4K CCD Imager) is an instrument
that has been developed and built at Steward Observatory thanks, in
part, to a Canadian Foundation for Innovation grant awarded to
G. Fontaine. E.M. Green at Steward has been the main driving force
behind this project. This partnership between the University of Arizona
and Universit\'e de Montr\'eal has been extremely fruitful and has allowed
us to gather high S/N light curves of pulsating hot subdwarf and white
dwarf stars. The instrument has also proved itself to be an excellent
imager in standard mode.

The CCD has a response of more than 95\% in the blue, ideal for studying
pulsating hot subdwarf and white dwarf stars. A wheel containing six 5''
$\times$ 5'' filters (currently $U$, $B$, $V$, $R$, $I$, and either
F555W or Schott 8612) can be used. The Mont4K is mounted exclusively at
the 1.55 m Kuiper Telescope on Mount Bigelow near Tucson. More details
about the instrument can be found at the
following web address,

\noindent http://james.as.arizona.edu/$\sim$psmith/61inch/CCD/CCDmanual.html

\begin{figure}[!ht]
\plotfiddle{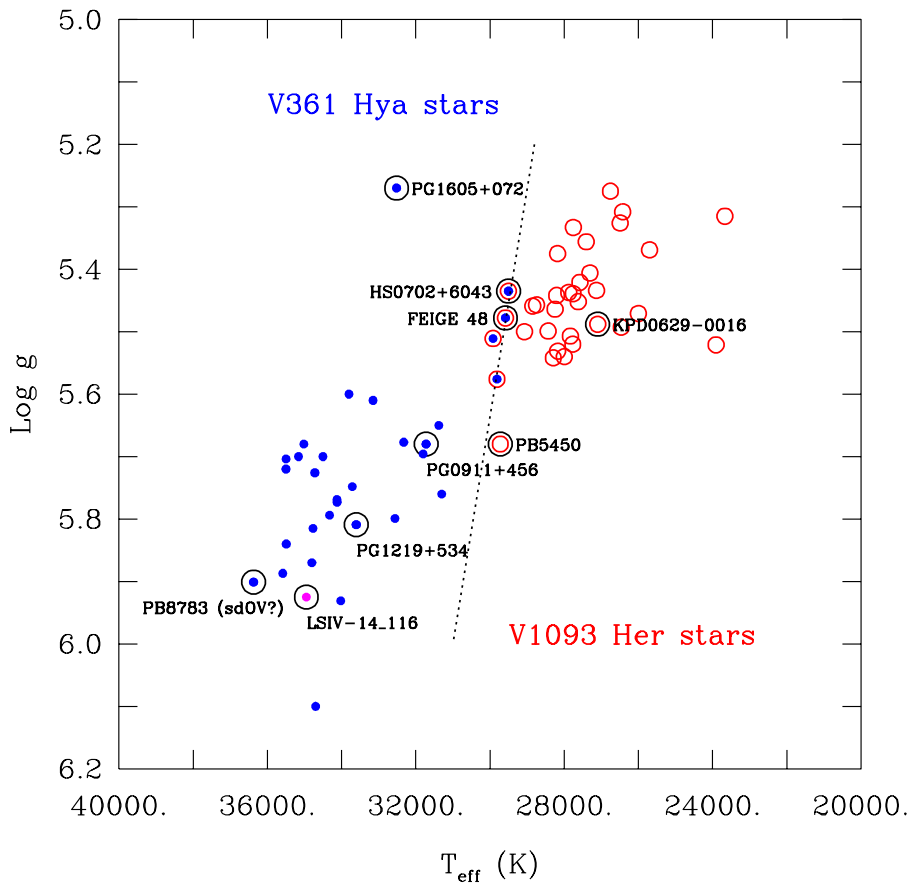}{12cm}{0}{65}{65}{-190}{-80}
\caption{Locations of pulsating hot subdwarfs in the effective
  temperature-surface gravity plane. The short-period pulsators of the
  V361 Hya type (sdBV$_{\rm r}$) are identified by the small filled circles,
  the long-period pulsators of the V1093 Her type (sdBV$_{\rm s}$) are
  identified by the small open circles, while variables of the hybrid
  type (sdBV$_{\rm rs}$) are identified by both symbols at the same
  time. The nine objects monitored so far through a Mont4K/Kuiper
  campaign are shown by the large open circles. Note that the PB5450
  campaign is scheduled for the fall of 2013. Note also that the
  position of PB8783 is quite uncertain; that star could be a much
  hotter sdOV pulsator.}  
\end{figure}

\section{The Campaigns}

We have taken advantage of the Arizona-Montr\'eal partnership to
periodically secure huge blocks of observing time on the 1.55 m Kuiper
Telescope, a privilege seldom granted on other telescopes of this
class. We selected targets deemed of particular interest, focussing
mostly on pulsating hot subdwarfs, but also including pulsating white
dwarfs. Although hugely successful, we have yet to exploit the results
of most of these campaigns because our collective attention over the
last few years has been diverted primarily on results obtained with the
$Kepler$ satellite.

\begin{table}[!ht]
\caption{Campaigns on Hot Subdwarfs with the Mont4K/Kuiper Combination}
\smallskip
\begin{center}
{\tiny
\begin{tabular}{ccccccccc}
\tableline
\noalign{\smallskip}
Dates & Target & $V$ & Type & Length & Duty cycle & Resolution &
Sampling time & Noise level\\
 & & & & (hr) & (\%) & ($\mu$Hz) & (s) & (\%)\\
\noalign{\smallskip}
\tableline
\noalign{\smallskip}
19/12/06-24/02/07 & PG0911+456 & 14.60 & sdBV$_{\rm r}$ & 57.1 & 3.5
& 0.17 & 38.30 & 0.0063 \\
\\
03/02/07-03/05/07 & PG1219+534 & 13.24 & sdBV$_{\rm r}$ & 198.7 & 9.3
& 0.13 & 19.44 & 0.0043 \\
\\
17/09/07-04/12/07 & PB8783 & 12.32 & sdBV$_{\rm r}$? & 182.5 & 9.8
& 0.15 & 21.20 & 0.0035 \\
 & & & sdOV? & & & & \\
01/11/07-14/03/08 & HS0702+6043 & 15.10 & sdBV$_{\rm rs}$ & 415.7 & 12.9
& 0.086 & 49.66 & 0.0044 \\
\\
01/01/09-17/05/09 & Feige 48 & 13.48 & sdBV$_{\rm rs}$ & 399.1 & 12.2
& 0.085 & 34.00 & 0.0035 \\
\\
10/11/09-10/03/10 & KPD0629$-$0016 & 14.91 & sdBV$_{\rm s}$ & $\sim$228 & ...
& ... & ... & ... \\
30/11/10-10/02/11 & '' & '' & '' & $\sim$203 & ...
& ... & ... & ... \\
15/09/10-02/11/10 & LSIV-14 116 & 13.0($B$) & He-sdBV & 53.5 & 4.6
& 0.24 & 51.78 & 0.0180 \\
\\
12/03/13-ongoing & PG1605+072 & 12.92 & sdBV$_{\rm r}$ & ... & ...
& ... & ... & ... \\
\\
fall 2013 & PB5450 & 13.06 & sdBV$_{\rm s}$ & ... & ...
& ... & ... & ... \\
\\
\noalign{\smallskip}
\tableline
\end{tabular}
}
\end{center}
\end{table}

The vital statistics of the campaigns carried out so far on pulsating
hot subdwarfs (including one in progress and another one planned for the
fall of 2013) are listed in Table 1. We use the nomenclature proposed by
\citet{K2010} to designate the various types of pulsators that were
considered. Note, in particular, the excellent temporal resolution
achieved and, above all, the rather remarkably low noise level
reached after our longest campaigns. This is the mean noise level
computed after prewhitening in the relevant frequency interval where
power was detected. This unique combination of resolution and
sensitivity from ground-based observations has allowed us to circumvent
most of the usual difficulties associated with short observing runs
carried out at a single site.

Figure 1 illustrates the locations of our hot subdwarf targets in
relation to other known pulsators in the effective temperature-surface
gravity domain. Along with three ``classic'' sdBV$_{\rm r}$ stars,
PG0911+456, PG1219+534, and PB8783 (but see below), we have also
dedicated long campaigns on the first-discovered hybrid (sdBV$_{\rm
rs}$) pulsator HS0702+6043 and on the long suspected pulsator of the
same kind, Feige 48. Furthermore, a campaign on the long-period
sdBV$_{\rm s}$ pulsator KPD0629$-$0016 was carried out in 2009-2010 to
back up a run that was obtained on the satellite $CoRoT$ in 2010 March
\citep{C2010}. Another similar campaign was carried out the
following season, but the data remain to be analyzed. Although the
difficulties of exploiting asteroseismology for long-period pulsating
hot subdwarfs observed from the ground have been well documented in
\citet{R2006}, it is hoped that the combination of space and ground
based data will help testing and improving the seismic model of
KPD0629$-$0016 proposed by \citet{V2010}. Also, we are hoping that the
difficulties mentioned in \citet{R2006} will be largely alleviated during
our next effort, planified for the fall of 2013, and dedicated to the
sdBV$_{\rm s}$ pulsator PB5450, since this object is the most compact of
its class (see Fig. 1) and, consequently, exhibits the shortest and more
easily characterized $g$-mode periods.

\begin{figure}[!ht]
\plotfiddle{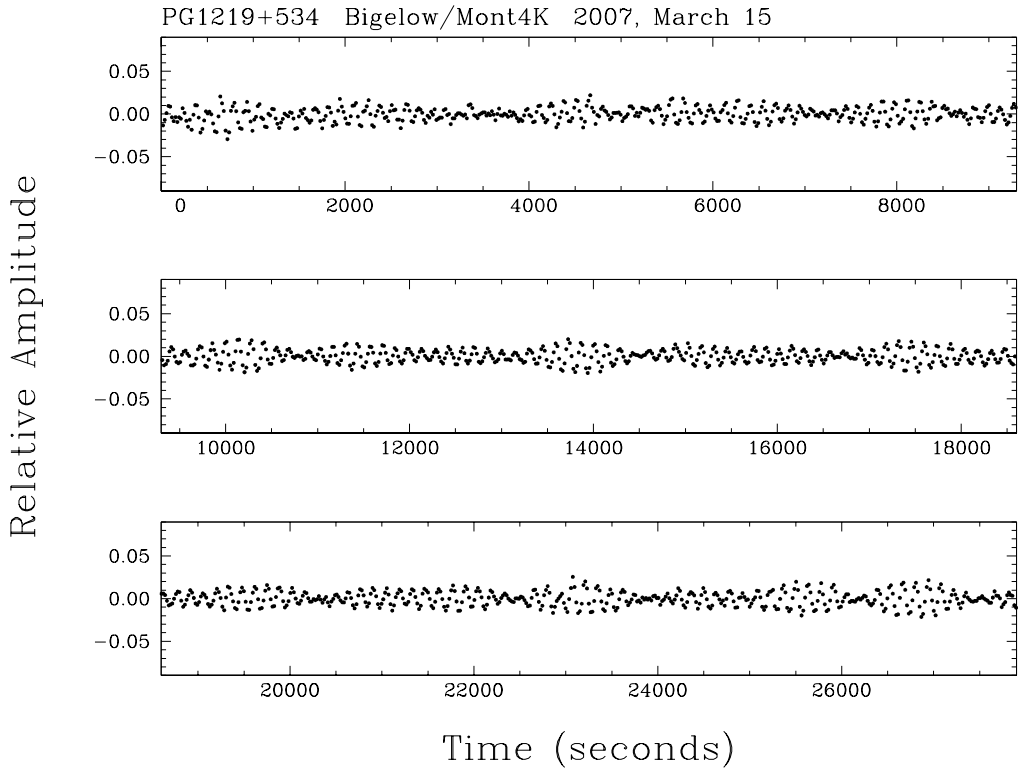}{10cm}{-90}{55}{55}{-205}{320}
\caption{Representative segment of the white light (CCD response +
  Schott 8612) light curve of the sdB$_{\rm r}$ star PG1219+534 obtained with
  the Mont4K/Kuiper combination.}
\end{figure}

\begin{figure}[!ht]
\plotfiddle{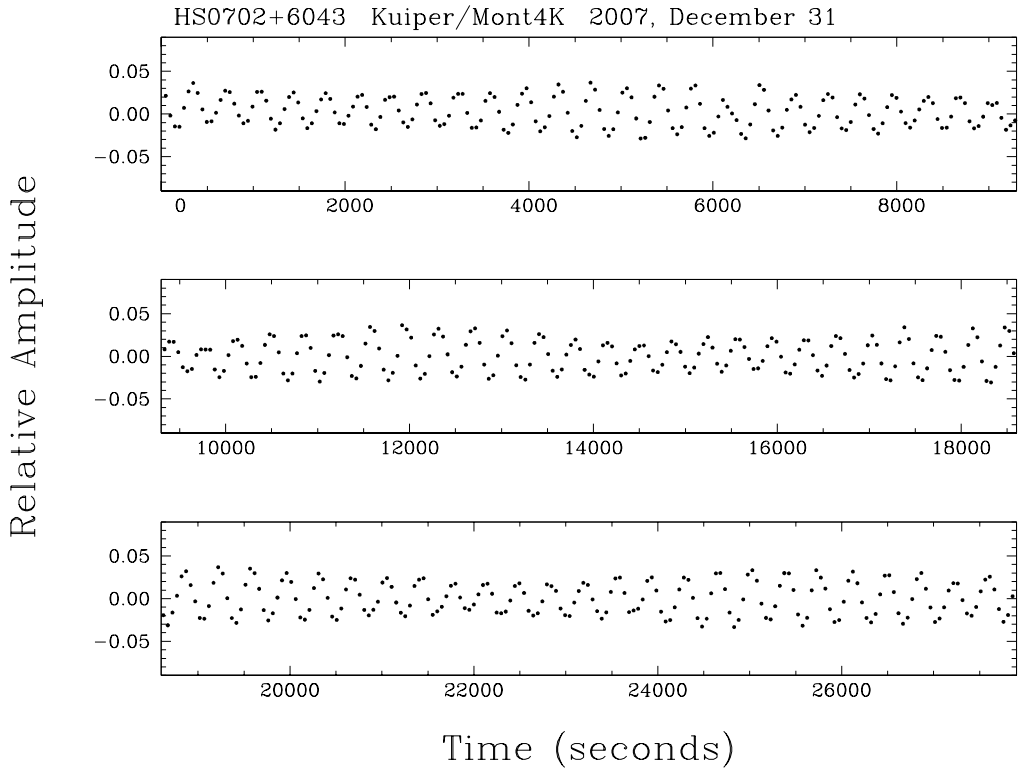}{10cm}{-90}{55}{55}{-205}{320}
\caption{Similar to Fig. 2, but for the hybrid sdB$_{\rm {rs}}$ star
  HS0702+6043.} 
\end{figure}

\begin{figure}[!ht]
\plotfiddle{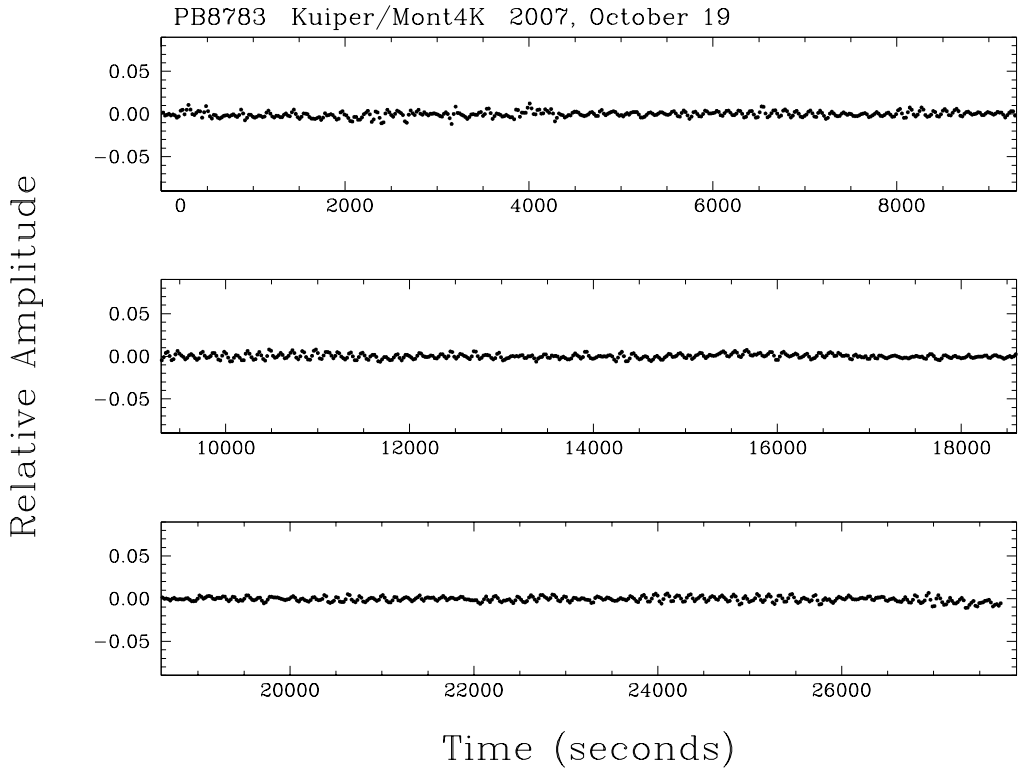}{10cm}{-90}{55}{55}{-205}{320}
\caption{Similar to Fig. 2, but for the pulsating (sdB or sdO?) star
PB8783.}
\end{figure}

\begin{figure}[!ht]
\plotfiddle{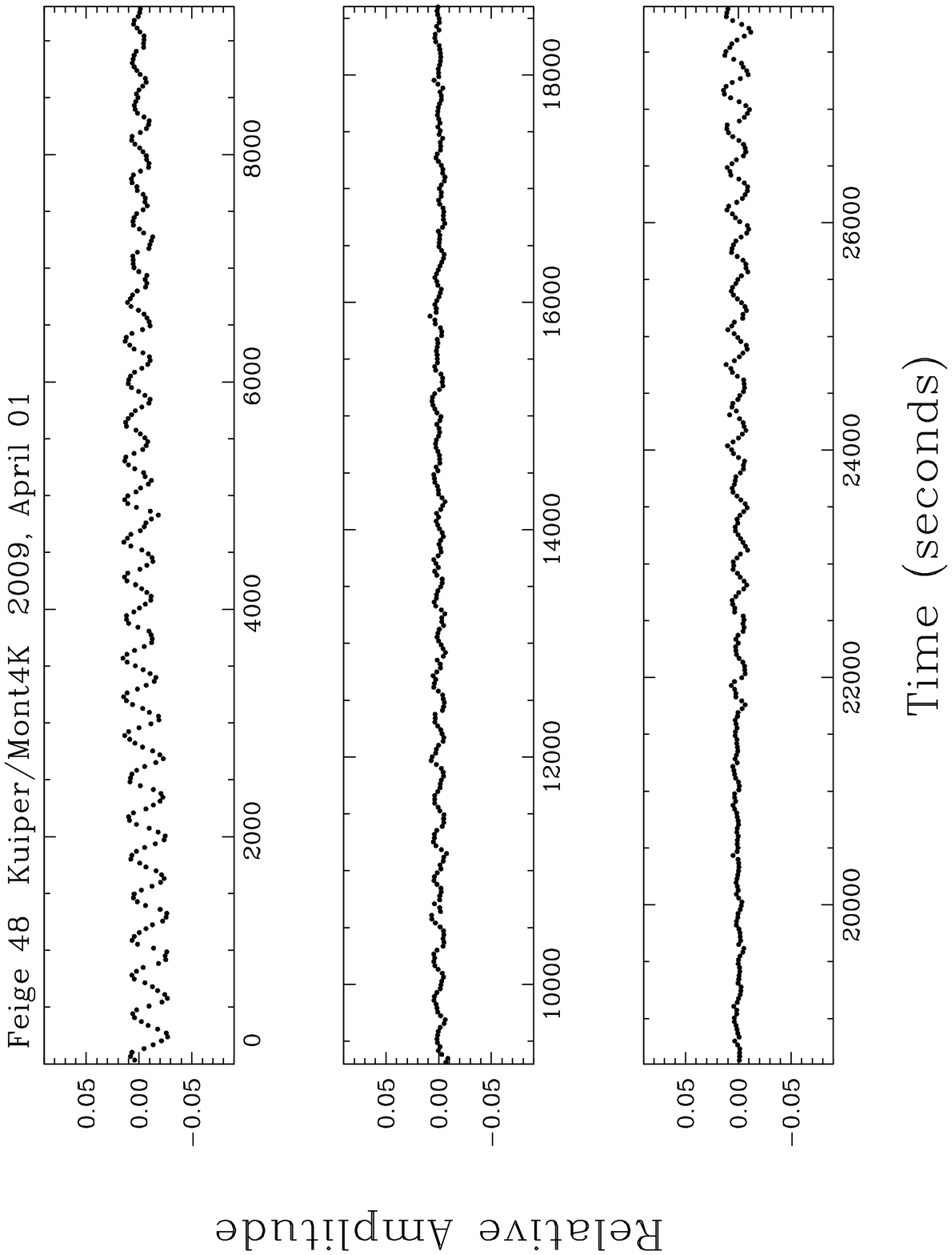}{10cm}{-90}{55}{55}{-205}{320}
\caption{Similar to Fig. 2, but for the hybrid sdB$_{\rm {rs}}$ star
  Feige 48.}
\end{figure}

In addition to the above, a relatively short campaign was carried out in
2010 on the unique and puzzling He-sdBV star LSIV-14 116
\citep{A2005,G2011}. Finally, a campaign is currently being
pursued on another challenging object, PG1605+072, possibly the most
intensely observed pulsating hot subdwarf star, but which has so far 
resisted detailed interpretation. The latest word on that target is
provided by \citet{L2012}.

In the remainder of this article, we provide some additional details on
four of our campaigns. In each case, we provide, on the exact same
scale, an example of a representative light curve obtained with the
Mont4K/Kuiper pair in order to facilitate the comparisons (Figs. 2, 3,
4, and 5).

\subsection{PG1219+534}

This target was selected in order to test, and ultimately improve, the
seismic model proposed by \citet{C2005}. That model had been developed
on the basis of the discovery of nine distinct pulsation modes during a
{\sl single} superlative night at the CFHT as reported by these
authors. Previous to that effort, only four modes were known for that
pulsator. We thus hoped that more modes still would be uncovered after
an extensive campaign with our setup on Mount Bigelow.

The campaign was somewhat disappointing in the sense that no new ($k$,$l$) 
modes were discovered beyond the nine $p$-modes already reported by 
\citet{C2005}. However, two of the nine redetected modes show obvious 
multiplet structure, best explained in terms of rotational splitting. In
particular, the second highest peak corresponding to a central period of
133.52 s in the Fourier transform appears to be made of a quintuplet. 
This mode stands out compared to the others as it shows relatively large
and complex amplitude and phase variations on a nightly basis. These
variations are well explained in terms of internal beating between the
components of the multiplet. Interpreted as rotational splitting, and
using the seismic model of \citet{C2005}, this leads to a preliminary
estimate of the (uniform) rotation period of 35.61 d for PG1219+534.

\subsection{HS0702+6043}

The first hybrid (sdBV$_{\rm rs}$) hot subdwarf pulsator was discovered
by \citet{S2006}. Although not particularly bright at $V$ = 15.10, its
light curve shows large amplitudes and the very nature of the object
makes it an intrinsically interesting target to study. The observations of
\citet{S2006} indicated that the light curve of HS0702+6043 is dominated
by a single mode, which made it an attractive target in the context of
the EXOTIME project to search for planets orbiting around pulsating hot
subdwarfs. The data that we gathered were used extensively by Ronnie
Lutz in his Ph.D. thesis. A recent summary of the EXOTIME efforts has
been presented by \citet{L2011}.

In our quest to ultimately exploit the seismic potential of HS0702+6043,
our campaign turned out to be quite promising. Indeed, our preliminary
reduction has led to the detection of a total of 13 $p$-modes and 10 
$g$-modes (+3 nonlinear peaks) with amplitudes larger than
4$\sigma$. Interestingly, no sign of rotational splitting is found,
despite a total baseline of 134.6 d. This suggests that HS0702+6043
rotates really slowly. 

\subsection{PB8783}

PB8783 is one of the original four hot subdwarfs that were discovered to
pulsate \citep{K1997}. Quite bright at $V$ = 12.32, \citet{O1998}
demonstrated early on its potential for seismology. They uncovered 11
pulsation modes after a two-week campaign. These modes belonged to 7
distinct complexes, including multiplets best explained in terms of
rotation. 

Despite rather small amplitudes (see Fig. 4), the light curve of PB8783
is rich with many oscillations. And indeed, our effort has led to the
detection of some 63 pulsation modes with amplitudes larger than
4$\sigma$ in that star. Contrary to HS0702+6043, however, PB8783 shows
clear evidence of rotational splitting as first suggested by
\citet{O1998}, and many of the detected modes are part of rotational
multiplets. In particular, we find a magnificent quintuplet, and an 
incomplete nontuplet. The latter may be yet the best observational
evidence in favor of a $l$ = 4 mode in a pulsating hot subdwarf. Using the
average spacing between the multiplet components, and in the absence of
a detailed seismic model for that star, we find a rotation timescale of
12.1 d for PB8783. 

It is not completely clear yet if PB8783 is a $\sim$36,000 K sdBV$_{\rm
r}$ star or a much hotter $\sim$50,000 K sdOV because its spectrum is
heavily polluted by the presence of a main sequence companion, and it is
consequently difficult to infer the atmospheric properties of the
pulsating hot subdwarf component of the binary system. \citet{O2012} has
recently argued that PB8783 is a hot sdO star, unrecognized as such
since the discovery of its pulsation in 1996. If correct, this implies
that PB8783 may be the first field equivalent of the pulsating sdO stars
discovered by \citet{R2011} in the globular cluster Omega Cen. Van
Grootel et al. (these proceedings) discuss the possible seismic
solutions for that star. 

\subsection{Feige 48}

The primary objective of this campaign is to test ultimately the seismic
model of Feige 48 proposed by \citet{V2008}. The latter is based on only
four multiplet structures associated with rotationally-split $p$-modes,
and it was hoped to detect many more modes from a long ground-based
effort at Mount Bigelow. In addition, we had secured $FUSE$ observing
time on Feige 48 to be gathered in early 2009 when we scheduled our
ground-based campaign. The idea was to assemble together as many
observational constraints as possible and, in the case of the
contemporary $FUSE$ observations, to exploit the amplitude-color
relation (FUV vs optical) in order to identify or constrain the $l$
index values for at least the dominant modes. Very unfortunately, the
$FUSE$ satellite ceased to operate in October 2008, just before the
beginning of our planned campaign.

The failed $FUSE$ component of our program was certainly a major
disappointment to us. However, our ground-based observational campaign
has been most successful on the front of uncovering new pulsations in
Feige 48, with the firm detection of 15
$g$-modes and 31 $p$-modes (+2 nonlinear peaks). The detection of
$g$-modes for the first time in that star makes it a hybrid pulsator. 
This discovery should not be too surprising in view of the location of
Feige 48 in the effective temperature-surface gravity plane (see, e.g.,
Fig. 1). We also find that rotational splitting is present, but it is
complicated due to the mixed character of the modes of interest.

An interesting bonus of our observations is the realization that Feige
48 is part of a reflection effect binary. Its companion is not a white
dwarf as previously believed \citep{O2004}, but a very cool 
main sequence star. Radial velocity measurements carried out by
E.M. Green at the MMT have led to a very precise determination of the
orbital period for the Feige 48 system: 8.2466210.00001 h (see the
paper by Latour et al. in these proceedings). We find a peak at that
period in our photometric data, thus proving that there is a reflection
effect in the system. Since such an effect cannot be caused by a tiny
white dwarf, the culprit must be a cool main sequence companion.

\acknowledgments{We are most grateful to all the people at Steward
  Observatory who have made the Mont4K instrument a reality. G.F. also
  wishes to acknowledge the essential contributions of the Canada
  Research Chair Program and of the Canadian Foundation for Innovation.}

\bibliographystyle{asp2010}
\bibliography{fontaine2.bbl}

\end{document}